\documentclass[traditabstract]{aa}
\usepackage{natbib}
\bibpunct{(}{)}{;}{a}{}{,} 
\usepackage{graphicx}
\usepackage{bm}
\usepackage{hyperref}
\usepackage{mathtools}
\usepackage{array}
\usepackage{tabularx}
\usepackage{multirow}
\usepackage{soul}
\usepackage{xcolor}
\usepackage{caption}
\usepackage{lineno}

\usepackage{txfonts}
\newcommand{\footnotemarkref}[1]{\textsuperscript{\ref{#1}}}

\begin{document}

\title{Constraints on AvERA Cosmologies \\
from Cosmic Chronometers and Type Ia Supernovae}

\author{
        Adrienn Pataki
        \inst{1}
        \fnmsep\thanks{corresponding author; \email{ patakia@student.elte.hu}}
    \and
        P\'eter Raffai
        \inst{1,2}
    \and
        Istv\'an Csabai
        \inst{3}
    \and
        G\'abor R\'acz
        \inst{4}
    \and
        Istv\'an Szapudi 
        \inst{5}      
        \fnmsep\thanks{MTA Guest Professor 2023}
        }

\institute{
        Institute of Physics and Astronomy, ELTE E\"otv\"os Lor\'and University, 1117 Budapest, Hungary
    \and
        HUN-REN–ELTE Extragalactic Astrophysics Research Group, 1117 Budapest, Hungary
    \and
        Department of Physics of Complex Systems, ELTE E\"otv\"os Lor\'and University, 1117 Budapest, Hungary
    \and
        Department of Physics, University of Helsinki, Gustaf H\"allstr\"omin katu 2, FI-00014 Helsinki, Finland
    \and
        Institute for Astronomy, University of Hawaii, 2680 Woodlawn Drive, Honolulu, HI 96822, USA
        }

\date{Received June 3, 2025; accepted June 20, 2025}

\abstract{We constrain AvERA cosmologies in comparison with the flat $\Lambda$CDM model using cosmic chronometer (CC) data and the Pantheon+ sample of type Ia supernovae (SNe Ia). The analysis includes fits to both CC and SN datasets using the \texttt{dynesty} dynamic nested sampling algorithm. For model comparison, we use the Bayesian model evidences and Anderson-Darling tests applied to the normalized residuals to assess consistency with a standard normal distribution. Best-fit parameters are derived within the redshift ranges $z \leq 2$ for CCs and $z \leq 2.3$ for SNe. For the baseline AvERA cosmology, we obtain best-fit values of the Hubble constant of ${H_0=68.32_{-3.27}^{+3.21}~\mathrm{km~s^{-1}~Mpc^{-1}}}$ from the CC analysis and ${H_0=71.99_{-1.03}^{+1.05}~\mathrm{km~s^{-1}~Mpc^{-1}}}$ from the SN analysis, each consistent within $1\sigma$ with the corresponding AvERA simulation value of $H(z=0)$. While both the CC and SN datasets yield higher Bayesian evidence for the flat $\Lambda$CDM model, they favor the AvERA cosmologies according to the Anderson-Darling test. We have identified signs of overfitting in each model, which suggests the possibility of overestimating the uncertainties in the Pantheon+ covariance matrix.}

\keywords{cosmological parameters --
          Cosmology: theory
          }

\maketitle
%

\section{Introduction}

The Hubble tension is the most notable discrepancy of modern cosmology. The expansion rate measured from late-time and early-time observations shows a $4-6\sigma$ deviation (\citet{Riess_2020}) if the standard $\Lambda$-cold dark matter ($\Lambda$CDM; see \citet{Peebles_Ratra_2003} for a review) cosmological model is used to compare the observations \citep{2023Univ....9...94H}. Measurements from the Cosmic Microwave Background (CMB) suggest a lower $H_0$ value \citep{Planck_2018}, while local universe observations, such as the Type Ia supernovae (\citet{Riess_et_al_2022}, \citet{Riess_2024}), indicate a higher expansion rate. This inconsistency challenges our understanding of the universe's expansion history and suggests potential new physics beyond the $\Lambda$CDM paradigm. To address this tension, alternative cosmological models such as the timescape and backreaction models have been proposed. The timescape model, introduced by \cite{2007PhRvL..99y1101W}  posits that the universe's inhomogeneous structure leads to differential aging in regions of varying gravitational potential. This model suggests that the observed acceleration of cosmic expansion could be attributed to time dilation effects in low-density voids, thereby reconciling the disparate $H_0$ measurements. Similarly, the backreaction models \citep{2000GReGr..32..105B, 2015CQGra..32u5021B} emphasize the influence of cosmic inhomogeneities on the universe's average expansion rate. By accounting for the effects of large-scale structures, such as voids and clusters, this model challenges the assumption of a perfectly homogeneous universe inherent in the $\Lambda$CDM framework. Incorporating backreaction effects can add late-time complexity to the expansion history, and it can adjust the effective Hubble constant, potentially resolving the observed tension between early and late universe measurements. These models offer promising avenues for resolving the Hubble tension, highlighting the need for a deeper understanding of cosmic inhomogeneities and their impact on the universe's expansion dynamics.

The Average Expansion Rate Approximation (AvERA) cosmology is a statistical, non-perturbative approach to model the cosmic backreaction \citep{Rácz_2017}. It builds on the general relativistic separate universe conjecture, which states that spherically symmetric regions in an isotropic universe behave as independent mini-universes with their own energy density -- a concept proven by Dai, Pajer \& Schmidt \citep{Dai_2015}. The AvERA approach uses these local densities to compute local expansion rates from the Friedmann equations and performs spatial averaging to estimate the overall expansion rate.

Local inhomogeneities influence the global expansion rate due to the nonlinearity of Einstein's equations \citep{2018CQGra..35xLT02B}. Under the AvERA approximation, the non-Gaussian distribution of matter and the growing low-density void regions cause the expansion to accelerate even with the flat $\Omega_{\Lambda}=0$ initial conditions. Consequently, this approach has the potential to eliminate the need for dark energy -- an unknown form of energy introduced within the concordance $\Lambda$CDM model -- to explain cosmic evolution. Furthermore, \citet{Rácz_2017} concluded that AvERA cosmology can resolve the Hubble tension.

The AvERA code is a collisionless $N$-body simulation that estimates the scale factor increment at each time step by averaging the local volumetric expansion rates of small subvolumes. Although local redshift evolution can vary across regions, the simulation uses a single global time step and applies a homogeneous rescaling of distances and velocities based on the effective scale factor, ensuring a one-to-one correspondence between time and redshift.

Within the AvERA framework, no closed-form analytic expression is available to describe the expansion history of the universe. Instead, each simulation outputs tabulated redshifts $z$ and Hubble parameters $H(z)$ at discrete time steps. The particle mass, corresponding to a coarse graining scale, is an adjustable parameter, that correlates with the matter density and thereby sets the expansion function, $E(z)$. According to \mbox{\citet{Rácz_2017}} the simulations were run with four different settings, using \mbox{$\{135,\,320,\,625,\,1080\}\times 10^3$} particles which correspond to \mbox{$\{9.4,\,3.96,\,2.03,\,1.17\}\times 10^{11}\,M_{\odot}$} coarse graining scales, within a comoving volume of \mbox{$147.62^3$ Mpc$^3$}. The overall expansion amplitude was fixed by anchoring the Hubble parameter at $z=9$ to \mbox{$H_{z=9}=1191.9 \,\textrm{km} \,\textrm{s}^{-1}\,\textrm{Mpc}^{-1}$}, complying with the Planck $\Lambda$CDM best-fit parameters -- thereby ensuring consistency with CMB measurements. Consequently, different coarse graining scale settings lead to different present-day Hubble constants, $H_{0,\textrm{model}}$ (listed in Table~\ref{tab:table3}).

In this paper, we use the four simulation outputs to test and constrain the AvERA cosmology with cosmic chronometers (CCs) and type~Ia supernovae (SNe~Ia). We probe \mbox{$E(z)=H(z)/H_{0,\mathrm{model}}$} evaluated at the data redshifts (by linear interpolation), while treating $H_0$ as a free parameter. This separation enables a shape-only analysis, independent of the Planck-calibration. Agreement between the fitted $H_0$ and $H_{0,\mathrm{model}}$ then indicates that AvERA, under that coarse graining setting, is simultaneously consistent with the dataset and the CMB.
In both tests, we compare the performance of the \mbox{AvERA} cosmologies to that of the flat $\Lambda$CDM model. Section \ref{sec:CC} and Section \ref{sec:SNIa} present the analyses and test results for CCs and SNe Ia, respectively, while Section \ref{sec:Conclusion} discusses our findings and summarizes our conclusions. Hereafter, we refer to AvERA cosmology as \mbox{"the AvERA model"}.

\section{Test with cosmic chronometers}
\label{sec:CC}

Under the assumption of the Friedmann-Lemaître-Robertson-Walker metric, the Hubble parameter is given by $$H(z)=-(1+z)^{-1}\,dz / dt.$$ The cosmic chronometer (CC) method, originally introduced by \citet{Jimenez_Loeb_2002} as the differential age method, estimates $dz/dt$ by measuring the age and redshift differences ($\Delta  t$ and $\Delta z$, respectively) between two ensembles of passively evolving massive galaxies that are separated by a small redshift interval. The dominant source of measurement uncertainty typically arises from the determination of $\Delta t$.

For our analysis, we used the most up-to-date compilation of $H(z)$ measurements obtained with the CC method \citep{Simon_et_al_2005,Stern_et_al_2010,Moresco_et_al_2012,Zhang_et_al_2014,Moresco_2015,Moresco_et_al_2016,Ratsimbazafy_et_al_2017,Borghi_et_al_2022,Tomasetti_2023} as listed in Table 1 of \citet{Moresco_2024}. Among the three flagged data points that should not be used jointly in an analysis -- because they are based on the same (or calibrated on the same) sample but obtained with different methods --, we retained only the one originally included in Table 1 of \citet{Moresco_et_al_2022}. We therefore used a total of 33 CC data points.

\begin{figure}
    \centering
 	\includegraphics[width=\columnwidth]{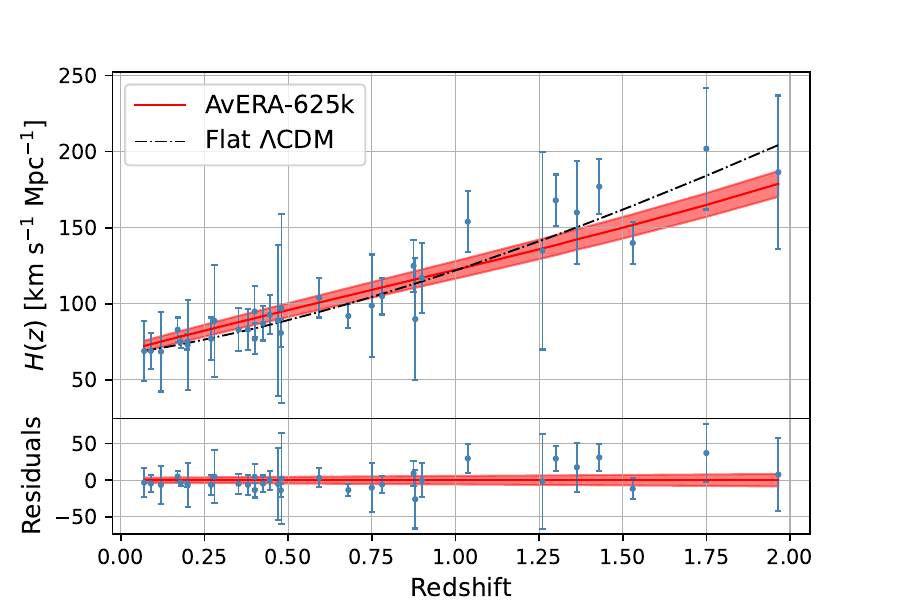}
     \caption{The $33$ CC data points from Table 1 of \citet{Moresco_2024} (blue dots with error bars), shown together with the best-fit $H(z)$ curves for the AvERA-625k model (red solid curve with $1\sigma$ confidence bands) and the flat $\Lambda$CDM model (black dash-dotted curve). The lower panel displays the residuals after subtracting the best-fit AvERA-625k $H(z)$ curve. The best-fit parameters for both models are given in Table~\ref{tab:table1}. "625k" indicates the particle number used in the AvERA simulation.}
     \label{fig:fig1}
\end{figure}

From the AvERA simulation outputs, we obtained $H(z_i)_{\rm sim}$ at the CC redshifts $z_i$ (with $i$ indexing the individual CC data points) by linear interpolation between the two adjacent tabulated values bracketing $z_i$. Although $H(z)_{\rm sim}$ appears nearly linear across the CC redshift range (see Figure \ref{fig:fig1}) and the simulation grid is reasonably dense, we estimated an upper bound on the interpolation error. This was quantified by comparing, at each tabulated redshift, the exact $H(z)_{\rm sim}$ value with that obtained by interpolating between its two neighboring $H(z)_{\rm sim}$ points. The relative deviation was calculated as $|H(z)_{\rm sim}/H(z)_{\rm sim, interp}-1|$. Across the CC redshift range, it remained below $0.12\%$ at maximum and below $0.03\%$ on average. We therefore conclude that linear interpolation provides sufficient accuracy for our analysis and thus does not affect our conclusions.

The expansion function at the $i$th CC redshift $z_i$ is then:
\begin{equation}\label{Eq:Ez_CC}
E(z_i)= \frac{H(z_i)_{\rm sim}}{H_{0 \textrm{,model}}},
\end{equation}
and the model $H(z)$ used for testing is:
\begin{equation}\label{eq:Hz_AvERA}
H(z_i)_{\rm model}=H_0 E(z_i).
\end{equation}

The $H(z)$ function for the flat $\Lambda$CDM model is given by:
\begin{equation}\label{Eq:Hz_LCDM}
H(z) = H_0E(z) = H_0 \sqrt{\Omega_\mathrm{m,0}\left( 1+z \right)^3+1-\Omega_\mathrm{m,0}},
\end{equation}
\noindent
where $\Omega_\mathrm{m,0}$, the present-day matter density parameter, and $H_0$ are free parameters. The contribution of radiation to the total density is neglected.

In our analysis, we fitted $H_0$ for the AvERA model, and $H_0$ and $\Omega_\mathrm{m,0}$ for the flat $\Lambda$CDM model, to the CC data using the \texttt{dynesty} ~\citep{Speagle_2020} \texttt{Python} package for dynamic nested sampling. We adopted uniform priors for all fits, with $H_0 \sim \mathcal{U}(50,80)\,\mathrm{km\,s^{-1}\,Mpc^{-1}}$ and $\Omega_\mathrm{m,0} \sim \mathcal{U}(0,1)$. The parameters were constrained  by minimizing
\begin{equation}\label{eq:chi}
\chi^2=\Delta \bm{D}^{T}C^{-1}_\mathrm{stat+syst} \Delta \bm{D},
\end{equation}
\noindent
where $\Delta \bm{D}$ is the vector of 33 residuals, with the $i$th component given by
\begin{equation}\label{eq:Di_CC}
\Delta D_i = H(z_i)_{\textrm{measured}} - H(z_i)_{\textrm{model}}
\end{equation}
\noindent
and $C_\mathrm{stat+syst}$ is the full statistical and systematic covariance matrix. Instructions on how to properly estimate the CC total covariance matrix and incorporate it into a cosmological analysis are provided in the \texttt{GitLab} repository\footnote{\label{note:emcee_code}\url{https://gitlab.com/mmoresco/CCcovariance}}, as referenced in \citet{Moresco_2024}.

\begin{table*}
\caption{\label{tab:table1} Model Fit and Test Results for CC Data}
\centering
\begin{tabular}{lccccc}
\hline \hline
 & AvERA-135k & AvERA-320k & AvERA-625k & AvERA-1080k & Flat $\Lambda$CDM \\
\hline
$H_0$ & $69.02_{-3.27}^{+3.27}$ & $68.67_{-3.28}^{+3.25}$ & $68.32_{-3.27}^{+3.21}$ & $67.86_{-3.23}^{+3.23}$ & $66.59_{-5.40}^{+5.26}$ \\
$\Omega_{\mathrm{m},0}$ & $0.332^*$ & $0.299^*$ & $0.279^*$ & $0.266^*$ & $0.336_{-0.063}^{+0.076}$ \\
\hline
$\chi^2$ & 19.17 & 19.32 & 19.14 & 18.94 & 14.59 \\
$\log_{10}\mathcal{B} \textrm{ (NS)}$ & 0.202 & 0.252 & 0.206 & 0.154 & 0 \\
$\log_{10}\mathcal{B} \textrm{ (AD)}$ & 0.013 & 0 (0.553) & 0.004 & 0.001 & 0.550 \\
\hline
\end{tabular}
\tablefoot{Best-fit parameter values and model comparison statistics for the CC dataset. The $H_0$ values are given in $[\mathrm{km\,s^{-1}\,Mpc^{-1}}]$ units. Simulated values of $\Omega_{\mathrm{m},0}$ values for the AvERA models (marked with asterisk) are also included for reference. The $\chi^2$ values were computed using the best-fit parameters. Bayes factors are reported for both the nested sampling evidence [$\log_{10} \mathcal{B}$ (NS)] and the Anderson-Darling test [$\log_{10} \mathcal{B}$ (AD)], as defined in the main text. While the nested sampling favors the flat $\Lambda$CDM model, the highest AD-test $p$-value is associated with the AvERA-320k model, corresponding to $\log_{10}p^{-1}_{\mathrm{max}}=0.553$ (shown in parentheses). However, both Bayes-factor measures indicate only weak preferences.}
\end{table*}

The 33 CC data points used for the fitting, along with their statistical errors, are shown in Figure~\ref{fig:fig1}. Figure~\ref{fig:fig1} also displays the best-fit $H(z)$ curves for the AvERA-625k model and the flat $\Lambda$CDM model, where "625k" refers to the particle number used in the AvERA simulation. The best-fit parameter values and their uncertainties, given as the 16th, 50th, and 84th percentiles of the parameter posteriors, are presented in Table~\ref{tab:table1}, while corner plots of the posteriors are available in our public code repository\footnote{\label{note:Zenodo_repo}\url{https://zenodo.org/records/15478706}}~\citep{Zenodo_repo}. Our best-fit $H_0$ and $\Omega_{\textrm{m,0}}$ values in the flat $\Lambda$CDM model are consistent with the results of \citet{Moresco_2023}, where ${H_0=66.7\pm 5.3~\mathrm{km~s^{-1}~Mpc^{-1}}}$ (within $0.015\sigma$) and $\Omega_{\textrm{m,0}}=0.33^{+0.08}_{-0.06}$ (within $0.06\sigma$).

In addition to parameter posterior estimation, the \texttt{dynesty} fit also yields the logarithm of the Bayesian evidence, $\log Z$, which quantifies the marginal likelihood of the data under each model. These values enable direct model comparison via the Bayes factor, defined as ${\log_{10}\mathcal{B}=\log_{10}(Z_\mathrm{max}/Z)}$, where $Z_\mathrm{max}$ is the highest evidence value among the models considered.
For model testing and comparison, we also adopt the Anderson-Darling (AD) test for normality~\citep{Anderson_Darling_1952,adtest_2024}, following the recommendations of \citet{Andrae_et_al_2010}. With this test, we evaluate a key property expected of a statistically valid model: that the fit residuals, once normalized by their observational uncertainties, are consistent with a standard normal distribution. A model is considered consistent with the null hypothesis of being the true model if the AD test yields a $p$-value of $p \geq 0.05$. For model comparison, we define Bayes factors from the AD test as ${\log_{10}\mathcal{B}=\log_{10}(p_\mathrm{max}/p)}$, where $p_\mathrm{max}$ is the highest $p$-value among the tested models. While $\log Z$ is based on the likelihood and thus on the $\chi^2$ statistics -- which compresses all residual deviations into a single scalar -- the AD test directly probes the full shape of the normalized residual distribution. This enables a more detailed statistical assessment of model performance and makes the AD test particularly sensitive to overfitting effects, as we will discuss in Section~\ref{sec:SNIa}. The $\chi^2$ values, along with the Bayes factors from both nested sampling and the AD test, are reported in Table~\ref{tab:table1}.

The model evidences derived from nested sampling indicate weak preferences for the flat $\Lambda$CDM model over all AvERA model (see Table \ref{tab:table1}). According to the AD test, both the AvERA and the flat $\Lambda$CDM models satisfy the ${p\geq 0.05}$ criterion, with the AvERA model being slightly favored by the CC data. The AD test results show no significant differences between the various AvERA models with different coarse graining scale settings. However, only the best-fit $H_0$ values for the \mbox{AvERA-320k} and AvERA-625k models are consistent within $1 \sigma$ with their corresponding simulated values of $H_{0 \textrm{,model}}$ (see Table~\ref{tab:table3}). This indicates that, for the 320k and 625k settings, the CC and CMB data are mutually consistent, whereas the CMB-calibrated 135k and 1080k models are incompatible with the CC dataset.

\section{Test with type Ia supernovae}
\label{sec:SNIa}

We tested and constrained the AvERA model and the flat $\Lambda$CDM model using the Pantheon+ sample\footnote{\label{note:Pantheon_data}\url{https://github.com/PantheonPlusSH0ES/DataRelease}} of SNe Ia~\citep{Scolnic_et_al_2022}, which comprises $1701$ light curves of $1550$ SNe within $z\lesssim 2.3$. Our methodology followed \citet{Brout_et_al_2022b}, where -- building on \citet{Tripp_1998} and \citet{Kessler_Scolnic_2017} -- the SN distance moduli were defined from the standardized SN brightnesses, using the parameters obtained from SALT2 light-curve fits~\citep{Guy_et_al_2007,Brout_et_al_2022a}:

\begin{equation}\label{eq:dist_mod}
\mu_\mathrm{SN}=m_B+\alpha x_1-\beta c - M_B -\delta_\mathrm{bias} + \delta_\mathrm{host},
\end{equation}
\noindent
where
\begin{equation}\label{eq:d_host}
\delta_\mathrm{host}=\gamma \times \left(e^{(\log_{10}M_\mathrm{*}-S)/\tau}\right)^{-1} -\frac{\gamma}{2}.
\end{equation}
Here, the $x_0$ amplitude, used as ${m_B\equiv -2.5\log_{10}(x_0)}$, the $x_1$ stretch, and the $c$ color parameters are from the light-curve fits. $\alpha$ and $\beta$ are global nuisance parameters, related to stretch and color, respectively. $M_B$ is the fiducial magnitude of an SN, and $\delta_\mathrm{bias}$ is a correction term to account for selection biases. The term $\delta_\mathrm{host}$ is the luminosity correction for the mass step \citep{Popovic_et_al_2021}, where $\gamma$ quantifies the magnitude of the luminosity difference between SNe in high- and low-mass galaxies; $M_*$ is the stellar mass of the host galaxy in units of $M_{\odot}$, while $S\sim 10$ (corresponding to $10^{10}\,M_{\odot}$) and $\tau$ define the step location and the step width, respectively. Altogether $m_B$, $x_1$, $c$, $\delta_\mathrm{bias}$ and $\log_{10}M_*$ are available for each SN in the Pantheon+ database~\footnotemarkref{note:Pantheon_data}, whereas the global nuisance parameters $\alpha$, $\beta$, $\gamma$, and $M_B$ must be fitted simultaneously with the cosmological parameters.

The model distance modulus is defined as
\begin{equation}\label{eq:dist_mod_model}
\mu(z)\equiv 5\log_{10}\left( \frac{d_L(z)}{10\ \mathrm{pc}} \right),
\end{equation}
\noindent
where $d_L(z)$ is the luminosity distance. The AvERA model does not provide explicit information about the present curvature density parameter, $\Omega_{k,0}$. However, within the redshift range of the Pantheon+ sample ($z \lesssim 2.3$), $d_L(z)$ has a weak dependence on the universe's spatial geometry. Moreover, since the Hubble parameter values $H(z)$ obtained from the AvERA simulations already encode the effects of curvature through the model's dynamics, the main curvature dependence is effectively incorporated into the integral for $d_L(z)$. Empirically, we also found that the explicit (geometrical) dependence of $d_L(z)$ on $\Omega_{k,0}$ has a negligible impact on the best-fit parameters. Therefore, we adopt the flat-universe approximation in our analysis, and leave the determination of $\Omega_{k,0}$ in the AvERA model for future work. The luminosity distance used in the fits, for both the AvERA and the flat $\Lambda$CDM models, is thus given by:
\begin{equation}\label{eq:dL}
d_L(z)=\left( 1+z \right)\frac{c}{H_0}\int_{0}^{z}\frac{\mathrm{d}z'}{E(z')}.
\end{equation}
In this expression, we used the cosmological redshift of each SN host galaxy in the CMB frame, corrected for peculiar velocity (denoted as $z_\mathrm{HD}$ in the Pantheon+ dataset~\footnotemarkref{note:Pantheon_data}; see \citealt{Carr_et_al_2022}).

For the AvERA model the expansion function $E(z)$ was computed as described in Section~\ref{sec:CC}. Specifically, we evaluated $E(z_i)$ at each SN redshift, $z_i$, using Equation~(\ref{Eq:Ez_CC}), where the corresponding $H(z_i)_{\rm sim}$ values were obtained by linear interpolation between adjacent $H(z)_{\rm sim}$ points in the simulation output. Across the SN redshift range, the maximum and average interpolation errors matched the CC values reported in Section~\ref{sec:CC}. In this model, only one cosmological parameter, $H_0$, was fitted simultaneously with the SN parameters. For the flat $\Lambda$CDM model, $E(z)$ is defined by Equation~(\ref{Eq:Hz_LCDM}), leading to two free cosmological parameters: $H_0$ and $\Omega_\mathrm{m,0}$.

Since the parameters $M_B$ and $H_0$ are degenerate (see Eqs.~(\ref{eq:dist_mod}) and~(\ref{eq:dist_mod_model}), (\ref{eq:dL})), we followed \citet{Brout_et_al_2022b} and, for the 77 SNe located in Cepheid hosts, we replaced the model distance modulus ($\mu (z)$) with the Cepheid-calibrated host-galaxy distance modulus ($\mu^{\textrm{Cepheid}}$) provided by SH0ES~\citep{Riess_et_al_2022}. Parameter estimation was performed using \texttt{dynesty} ~\citep{Speagle_2020}, minimizing $\chi^2$ as defined in Equation (\ref{eq:chi}), with

\begin{equation}\label{eq:Di_SN}
\Delta D_i = \begin{cases}
    \mu_{\textrm{SN,}i}-\mu_i^{\textrm{Cepheid}} & \text{$i \in$ Cepheid host}\\
    \mu_{\textrm{SN,}i}-\mu_i(z) & \text{otherwise}
    \end{cases}
\end{equation}
\noindent
representing the $i$th component of the SN residual vector.
The full statistical and systematic covariance matrix, $C_\mathrm{stat+syst}$, from \citet{Brout_et_al_2022b}~\footnotemarkref{note:Pantheon_data} was used in the fits and already accounts for the uncertainties in Cepheid-calibrated host distances. Uniform priors were applied for all parameters, with ranges listed in Table \ref{tab:priors}.

\begin{table}[ht]
\caption{\label{tab:priors} Uniform Prior Ranges Used in the SN Fit}
\centering
\begin{tabular}{clcl}
\hline \hline
Parameter & Prior range & Parameter & Prior range \\
\hline
$H_0$ & $\mathcal{U}(62,\,78)$ & $\alpha$ & $\mathcal{U}(0,\,0.2)$ \\
$M_B$ & $\mathcal{U}(-20,\,-18.8)$ & $\beta$ & $\mathcal{U}(2.5,\,3.5)$ \\
$\Omega_\mathrm{m,0}$ & $\mathcal{U}(0,\,1)$ & $\gamma$ & $\mathcal{U}(-0.1,\,0.1)$ \\
\hline
\end{tabular}
\tablefoot{The $H_0$ ranges are given in $[\mathrm{km\,s^{-1}\,Mpc^{-1}}]$.}
\end{table}

In the determination of best-fit model parameters, we followed the approach of \citet{Amanullah_et_al_2010}, \citet{Riess_et_al_2022}, and others, by employing the sigma clipping technique during model fitting, iteratively removing data points that deviated more than $3\sigma$ from the global fits until no such outliers remained. This method, as demonstrated by simulations such as those from \citet{Kowalski_et_al_2008}, helps preserve the fit results in the absence of contamination while mitigating the influence of any outliers. For the $1701$ SN data points, sigma clipping eliminated ${N=\left\{ 15,16,16,16 \right\}}$ for the AvERA model with ${\left\{ 135\mathrm{k}, 320\mathrm{k}, 625\mathrm{k}, 1080\mathrm{k} \right\}}$ particles, and $N=15$ for the flat $\Lambda$CDM model.

\begin{figure}
    \centering
 	\includegraphics[width=\columnwidth]{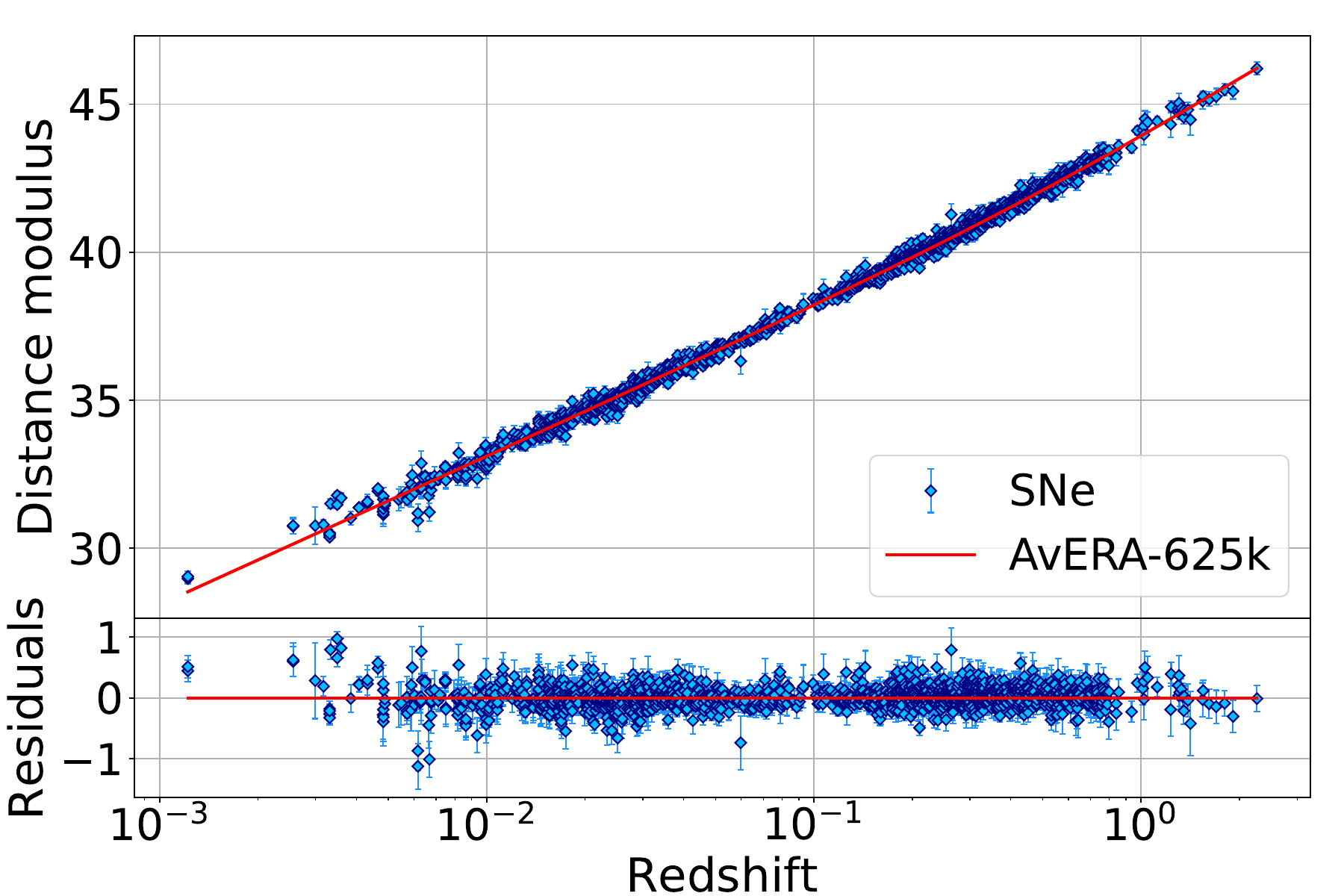}
     \caption{Distance moduli for 1701 SN Ia observations (blue diamonds with error bars) and the AvERA-625k model (red solid curve), computed using the best-fit parameters (see Table~\ref{tab:table2}). SN data are from the Pantheon+ sample~\citep{Scolnic_et_al_2022}. The lower panel shows the residuals, obtained by subtracting the model $\mu (z)$ curve.}
     \label{fig:fig2}
\end{figure}

We present the best-fit parameters and their uncertainties, given as the 16th, 50th, and 84th percentiles of the parameter posteriors, along with the test statistics in Table~\ref{tab:table2}. The posterior distribution plots are available in our public code repository~\footnotemarkref{note:Zenodo_repo}~\citep{Zenodo_repo}. Figure~\ref{fig:fig2} shows the distance moduli for all 1701 SNe, computed using the best-fit SN parameters of the AvERA-625k model, together with the corresponding $\mu (z)$ curve based on the best-fit value of $H_0=71.99_{-1.03}^{+1.05}~\mathrm{km~s^{-1}~Mpc^{-1}}$. As shown in Table~\ref{tab:table3}, only the AvERA-625k model exhibits agreement within $1 \sigma$ between the best-fit $H_0$ and the simulated value $H_{0 \textrm{,model}}$, indicating that SN and CMB data are mutually consistent only under the 625k setting.

\begin{figure}[h!]
    \centering
 	\includegraphics[width=\columnwidth]{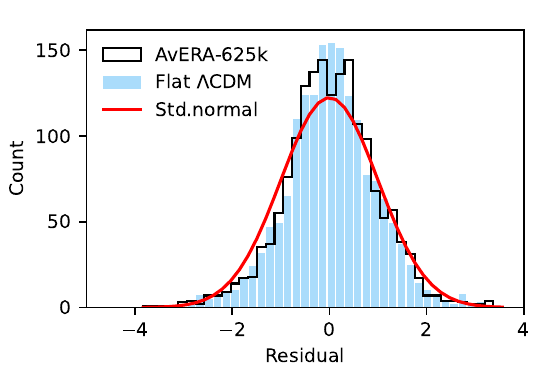}
    \includegraphics[width=\columnwidth]{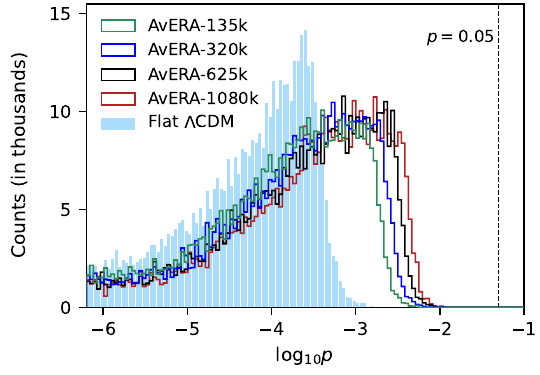}
     \caption{Upper panel: Histograms of normalized residuals for the best-fit AvERA-625k and flat $\Lambda$CDM models, fitted to the Pantheon+ SN Ia data~\citep{Scolnic_et_al_2022}. The red curve represents the standard normal distribution expected for the true model underlying the data. Lower panel: Histograms of AD-test ${\log_{10}p}$ values for half a million realizations sampled from the joint posterior distributions. The dashed vertical line marks the significance threshold of $p = 0.05$ (${\log_{10}p=-1.301}$) for consistency with the null hypothesis of being the true model. While none of the realizations reach this level, the preference for the AvERA model is clearly evident.}
     \label{fig:fig3}
\end{figure}

\begin{table*}
\caption{\label{tab:table2} Model Fit and Test Results for SN Ia Data}
\centering
\begin{tabular}{lccccc}
\hline \hline
 & AvERA-135k & AvERA-320k & AvERA-625k & AvERA-1080k & Flat $\Lambda$CDM \\
\hline
$H_0$ & $72.16_{-1.01}^{+1.04}$ & $72.06_{-1.01}^{+1.05}$ & $71.99_{-1.03}^{+1.05}$ & $71.94_{-1.02}^{+1.03}$ & $73.13_{-1.05}^{+1.06}$ \\
$M_B$ & $-19.201_{-0.031}^{+0.031}$ & $-19.200_{-0.031}^{+0.031}$ & $-19.200_{-0.030}^{+0.031}$ & $-19.198_{-0.031}^{+0.030}$ & $-19.210_{-0.031}^{+0.031}$ \\
$\alpha$ & $0.147_{-0.004}^{+0.004}$ & $0.146_{-0.004}^{+0.004}$ & $0.146_{-0.004}^{+0.004}$ & $0.147_{-0.004}^{+0.004}$ & $0.149_{-0.004}^{+0.004}$ \\
$\beta$ & $2.942_{-0.073}^{+0.072}$ & $2.941_{-0.071}^{+0.075}$ & $2.937_{-0.072}^{+0.073}$ & $2.951_{-0.073}^{+0.073}$ & $2.981_{-0.073}^{+0.073}$ \\
$\gamma$ & $0.010_{-0.011}^{+0.011}$ & $0.011_{-0.011}^{+0.011}$ & $0.010_{-0.011}^{+0.011}$ & $0.009_{-0.011}^{+0.011}$ & $0.010_{-0.011}^{+0.011}$ \\
$\Omega_{\mathrm{m},0}$ & $0.332^*$ & $0.299^*$ & $0.279^*$ & $0.266^*$ & $0.328_{-0.018}^{+0.018}$ \\
\hline
$\chi^2$ & 1539 & 1543 & 1547 & 1551 & 1521 \\
$\log_{10}\mathcal{B} \textrm{ (NS)}$ & 2.775 & 3.649 & 4.603 & 5.495 & 0 \\
$\log_{10}\mathcal{B} \textrm{ (AD)}$ & 0.223 & 0.148 & 0.074 & 0 (2.782) & 0.761 \\
\hline
\end{tabular}
\tablefoot{Best-fit parameter values and model comparison statistics for the SN dataset. The $H_0$ values are given in $[\mathrm{km\,s^{-1}\,Mpc^{-1}}]$ units. Simulated values of $\Omega_{\mathrm{m},0}$ values for the AvERA models (marked with asterisk) are also included for reference. The $\chi^2$ values were computed using the best-fit parameters. Bayes factors are shown for both the nested sampling evidence [$\log_{10} \mathcal{B}$ (NS)] and the Anderson-Darling test [$\log_{10} \mathcal{B}$ (AD)], as defined in the main text in Section \ref{sec:CC}. While none of the models satisfy the AD test's $p\geq 0.05$ criterion, the test favors the AvERA-1080k model (with $\log_{10}p^{-1}_{\mathrm{max}}=2.782$) and indicates a clear preference for the AvERA model over the flat $\Lambda$CDM model. This contrasts with the nested sampling Bayes factors, which strongly favor the flat $\Lambda$CDM model. The discrepancy is due to the flat $\Lambda$CDM model overfitting the SN data (see Section \ref{sec:SNIa} for a detailed discussion).}
\end{table*}

Our best-fit $H_0$ and $\Omega_{\mathrm{m},0}$ values in the flat $\Lambda$CDM model are consistent with the results of \citet{Brout_et_al_2022b}, where ${H_0=73.6\pm 1.1~\mathrm{km~s^{-1}~Mpc^{-1}}}$ (within $0.3\sigma$) and ${\Omega_{\mathrm{m},0}=0.334\pm 0.018}$ (within $0.2\sigma$). Our $H_0$ is also consistent with ${H_0=73.30\pm 1.04~\mathrm{km~s^{-1}~Mpc^{-1}}}$ (within $0.1\sigma$) and ${H_0=73.04\pm 1.04~\mathrm{km~s^{-1}~Mpc^{-1}}}$ (within $0.06\sigma$), obtained by \citet{Riess_et_al_2022} with and without the inclusion of high-redshift ($z\in[0.15,0.8)$) SNe, respectively. However our $H_0$ in the flat $\Lambda$CDM model is in $4.8\sigma$ tension with the Planck+BAO ${H_0=67.66\pm 0.42~\mathrm{km~s^{-1}~Mpc^{-1}}}$ \citep{Planck_2018}, confirming the Hubble tension.

For model comparison, we applied the same methods as in Section~\ref{sec:CC}: model evidences and AD test to the residuals. The normalized residuals were computed as $\bm{r}= L^{-1}\Delta \bm{D}$, with $C_\mathrm{stat+syst}=LL^T$ the Cholesky factorization of the covariance. The $\chi^2$ values, along with the Bayes factors derived from nested sampling and from the AD test, are reported in Table~\ref{tab:table2}. Since the nested sampling evidence ($\log Z$) is generated during the fitting process, we report $\log_{10}\mathcal{B}$ (NS) values from the initial fit iteration, which used the complete SN dataset prior to any sigma clipping, in order to maintain consistency across all statistics.

Although none of the best-fit models satisfied the $p\geq 0.05$ criterion for consistency with the SN data, the AD test-based Bayes factors indicate that the \mbox{AvERA} model is at least substantially favored over the flat $\Lambda$CDM model. As shown in Table~\ref{tab:table2}, this ranking is in contrast to that implied by the Bayes factors from nested sampling, which strongly favors the flat $\Lambda$CDM model. This discrepancy is due to the flat $\Lambda$CDM model overfitting the SN data, as illustrated in Figure~\ref{fig:fig3}. The AD test is particularly sensitive to such effects, as it responds to subtle deviations from the expected normal distribution of residuals, which may not be captured by global statistics like $\chi^2$ or $\log Z$. In the upper panel of Figure~\ref{fig:fig3}, the histogram of normalized residuals is shown, which are expected to follow a standard normal distribution for the true model underlying the SN data. Both the flat $\Lambda$CDM and AvERA-625k models yield residuals that are overly concentrated near zero (i.e., under–dispersed relative to $\mathcal{N}(0,1)$), with sample standard deviations of $\sigma = 0.93$ and $\sigma = 0.94$, respectively. The lower panel of Figure~\ref{fig:fig3} displays histograms of AD-test ${\log_{10}p}$ values computed for half a million samples drawn from the joint posterior distributions of the SN fit. While none of the flat $\Lambda$CDM or AvERA realizations satisfy the $p\geq 0.05$ ($\log_{10}p\geq -1.3$) criterion, the preference for the AvERA model remains clearly evident. The highest (median) $p$-value for the flat $\Lambda$CDM model is exceeded by $\left\{0.2\%,\,0.7\%,\,2.6\%,\,6.1\%\right\}$ ($\left\{64\%,\,69\%,\,71\%,\,73\%\right\}$) of the $p$-values for the AvERA model with ${\left\{ 135\mathrm{k}, 320\mathrm{k}, 625\mathrm{k}, 1080\mathrm{k} \right\}}$ particles.

Despite its lower $\chi^2$ value, the flat $\Lambda$CDM model is inconsistent with the Pantheon+ data according to the AD test. This is due to the model overfitting the SN data, potentially as a consequence of the error estimation strategy \citep{Raffai_2025}. While conservative error estimation is generally a sound approach for robust parameter inference, it can result in the tested models overfitting data, potentially leading to false conclusions in model comparisons. Consistent with our finding, \citet{Keeley_2024} also concluded that the flat $\Lambda$CDM model overfits the Pantheon+ SN data, hinting that SN measurement errors may have been conservatively overestimated. Another potential source of overfitting is that the preprocessed observational data may not be entirely independent from the concordance (flat $\Lambda$CDM) cosmological model. For example, as noted by \citet{Carr_et_al_2022}, $z_{\textrm{HD}}$ values in the Pantheon+ dataset are corrected for peculiar velocities of SN host galaxies calculated by assuming flat $\Lambda$CDM cosmology. This model dependency may contribute to the flat $\Lambda$CDM model overfitting the SN data, although \citet{Carr_et_al_2022} note that the cosmology-dependence of the correction is weak.

\section{Conclusions}
\label{sec:Conclusion}

In Sections~\ref{sec:CC}--\ref{sec:SNIa}, we tested and compared the predictions of the AvERA model with those of the flat $\Lambda$CDM model. A comprehensive summary of our results is provided in Tables~\ref{tab:table1} and \ref{tab:table2}.

\begin{table*}
\caption{\label{tab:table3} Deviations in $H_0$ between Simulation and Test}
\centering
\begin{tabular}{lccccc}
\hline \hline
 & $H_{0 \textrm{,model}}$ & $H_0$ CC test & deviation CC & $H_0$ SN test & deviation SN \\
\hline
AvERA-135k & 65.39 & $69.02_{-3.27}^{+3.27}$ & 1.11$\sigma$ & $72.16_{-1.01}^{+1.04}$ & 6.68$\sigma$ \\
AvERA-320k & 68.91 & $68.67_{-3.28}^{+3.25}$ & 0.07$\sigma$ & $72.06_{-1.01}^{+1.05}$ & 3.10$\sigma$ \\
AvERA-625k & 71.38 & $68.32_{-3.27}^{+3.21}$ & 0.95$\sigma$ & $71.99_{-1.03}^{+1.05}$ & 0.59$\sigma$ \\
AvERA-1080k & 73.14 & $67.86_{-3.23}^{+3.23}$ & 1.63$\sigma$ & $71.94_{-1.02}^{+1.03}$ & 1.17$\sigma$ \\
\hline
\end{tabular}
\tablefoot{$H_0$ values are given in $[\mathrm{km}\ \mathrm{s}^{-1}\ \mathrm{Mpc}^{-1}]$ units. The deviations indicate the tension between the best-fit $H_0$ values obtained from the CC and SN fits and the corresponding simulated $H_{0 \textrm{,model}}$ values \citep{Rácz_2017} for each AvERA setting. While the CC test does not strongly favor any specific coarse graining scale, the SN test prefers higher particle numbers. Among the tested AvERA models, the 625k simulation shows the best agreement, with both CC and SN best-fit values lying within $1\sigma$ of the simulated $H_{0 \textrm{,model}}$.}
\end{table*}

The CC and SN datasets favor the AvERA model over the flat $\Lambda$CDM model according to the AD test, but indicate a preference for the flat $\Lambda$CDM model based on the nested sampling evidences. Among the tested AvERA models, the 625k setting stands out: the simulated $H_{0,\mathrm{model}}$ lies within the $1\sigma$ credible interval of the fitted $H_0$ in both the CC and SN analyses (Table~\ref{tab:table3}). Since $H_{0,\mathrm{model}}$ results from calibrating the simulation to the CMB, this concordance implies mutual consistency of the CC, SN, and CMB data under the AvERA-625k model; accordingly, we adopt it as our baseline AvERA model. Within this model, such concordance points to a potential alleviation of the Hubble tension -- that is, the discrepancy between CMB- and SN-based determinations of $H_0$.

Only four outputs, corresponding to four discrete coarse graining settings, were available for this analysis; however, the true scale could in principle lie between them. To seek the optimal coarse graining scale, we performed a joint CC+SN fit for the AvERA model, treating the coarse graining scale as a free parameter. To achieve this, we applied linear interpolation between the four simulation outputs to compute the model $H(z)$ for any integer particle number within the range [135k, 1080k]. In this analysis, the SN parameters were fixed to the average of their best-fit values from Table~\ref{tab:table2}, as the table shows that these parameters vary only minimally across different graining scales. The implementation code and the corresponding corner plot are available in our code repository~\footnotemarkref{note:Zenodo_repo}~\citep{Zenodo_repo}. This analysis did not reveal a single preferred graining scale, as indicated by a nearly uniform posterior distribution of the particle number. We obtained ${H_0=71.64^{+4.43}_{-4.55}~\mathrm{km~s^{-1}~Mpc^{-1}}}$, which is in only $0.06\sigma$ tension with the simulated value of $H_0$ for the AvERA-625k model, further supporting its choice as the baseline AvERA model.

\begin{acknowledgements}
The authors would like to thank Bence Bécsy for his assistance with \texttt{dynesty}. This project has received funding from the HUN-REN Hungarian Research Network and was also supported by the NKFIH excellence grant TKP2021-NKTA-64 and  NKFI-147550. IS acknowledges NASA grants 80NSSC24K1489 and 24-ADAP24-0074 and thanks the hospitality of the MTA-CSFK Lend\"ulet "Momentum" Large-Scale Structure (LSS) Research Group at Konkoly Observatory supported by a \emph{Lend\"ulet} excellence grant by the Hungarian Academy of Sciences (MTA). GR acknowledges the support of the Research Council of Finland grant 354905 and the support by the European Research Council via ERC Consolidator grant KETJU (no. 818930). 
\end{acknowledgements}

\vspace{5mm}

\bibliographystyle{aa}
\bibliography{AvERA_Test}{}

\end{document}